\documentclass[journal]{IEEEtran}
\ifCLASSINFOpdf
\else
\fi
\def\ps@headings{%

\def\@oddhead{\mbox{}\scriptsize\rightmark \hfil \thepage}%

\def\@evenhead{\scriptsize\thepage \hfil \leftmark\mbox{}}%

\def\@oddfoot{}%

\def\@evenfoot{}}

\makeatother

\usepackage{multirow}
\usepackage{amsfonts}
\usepackage{amsmath}
\usepackage{graphicx}
\usepackage{algorithm}

\usepackage{setspace}
\usepackage{subfigure}
\usepackage{array}
\usepackage{overpic}
\usepackage{amssymb}
\newcommand{\PreserveBackslash}[1]{\let\temp=\\#1\let\\=\temp}

\newcolumntype{C}[1]{>{\PreserveBackslash\centering}p{#1}}
\newcolumntype{R}[1]{>{\PreserveBackslash\raggedleft}p{#1}}
\newcolumntype{L}[1]{>{\PreserveBackslash\raggedright}p{#1}}
\begin{document}

\title{Joint Data Scheduling and FEC Coding for Multihomed Wireless Video Delivery}

\author{Jasmin Fantel$^{\ast}$ and Yan~Gao$^{\star}$\\
$^{\ast}$Telecommunication Technology Research Center, University of Innstingbruck,\\
$^{\star}$School of Sciences, Southwest University of Science and Technology,\\

}



%


\maketitle

\begin{abstract}
This paper studies the problem of mobile video delivery in heterogenous wireless networks from a server to multihomed device. Most existing works only consider delivering video streaming on single path which bandwidth is limited causing ultimate video transmission rate. To solve this live video streaming transmission bottleneck problem, we propose a novel solution named Joint Data Allocation and Fountain Coding (JDAFC) method that contain below characters: (1) path selection, (2) dynamic data allocation, and (3) fountain coding. We evaluate the performance of JDAFC by simulation experiments using Exata and JVSM and compare it with some reference solutions. Experimental results represent that JDAFC outperforms the competing solutions in improving the video peak signal-to-noise ratio as well as reducing the end-to-end delay.


\end{abstract}
\begin{keywords}
mobile video delivery; heterogeneous wireless networks; multi-homing; fountain code
\end{keywords}

\IEEEpeerreviewmaketitle

\section{Introduction}
Recent years have witnessed the proliferation of various mobile devices (e.g., smartphone and pad). Consequently, mobile video streaming (e.g., Hollywood [1] and Metacafe [2]) becoming the most important part of network flow. We can draw a conclusion from the Cisco Visual Index [3] that the proportion of mobile video streaming is $57\%$ in 2014 and will increase to $69\%$ by the year 2019. Therefore, it is significant to study on mobile video quality guarantee.

With the advancements in wireless communication technology, various network access technology had been developed (e.g., LTE, WLAN and WiMAX), but there are still many problems due to the bandwidth limitation, consequently video quality is hard to promotion. LTE, WLAN and WiMAX are not applied widely and bottleneck arises when users increasing because of bandwidth scarcity [4-5]. Meanwhile the cover range of WLAN is finite and the quality of signal is unstable causing WLAN could not guarantee video streaming quality. As a result, the limitation of single wireless network turn research attentions to multi-path video transmission.

To solve the channel loss problem, researchers proposed joint source-channel coding (JSCC) and proved this method was effective. But the bandwidth limitation is still a critical problem. Contemporary JSCC approaches (e.g., [7-8]) are focus on single path to transfer mobile video streaming. So we proposed a multi-path mobile video streaming transfer solution to settle this problem. But the problem becomes more complicated when considering heterogenous wireless network. As shown in Fig. 1, at location 1 the cellular link is unstable and results in  video stalls that degrade the user experience. At location 2 the user switch to WLAN network and request to the video server, but WLAN could not guarantee the video transfer quality and the situation turns even worse. And the video quality become better when user connect to WiMAX at location 3. We can arrive at the conclusion that the user-perceived video quality will only be reduced when involving an unreliable wireless access network.

\begin{figure}
\centering
 \includegraphics[width=0.45\textwidth,keepaspectratio]{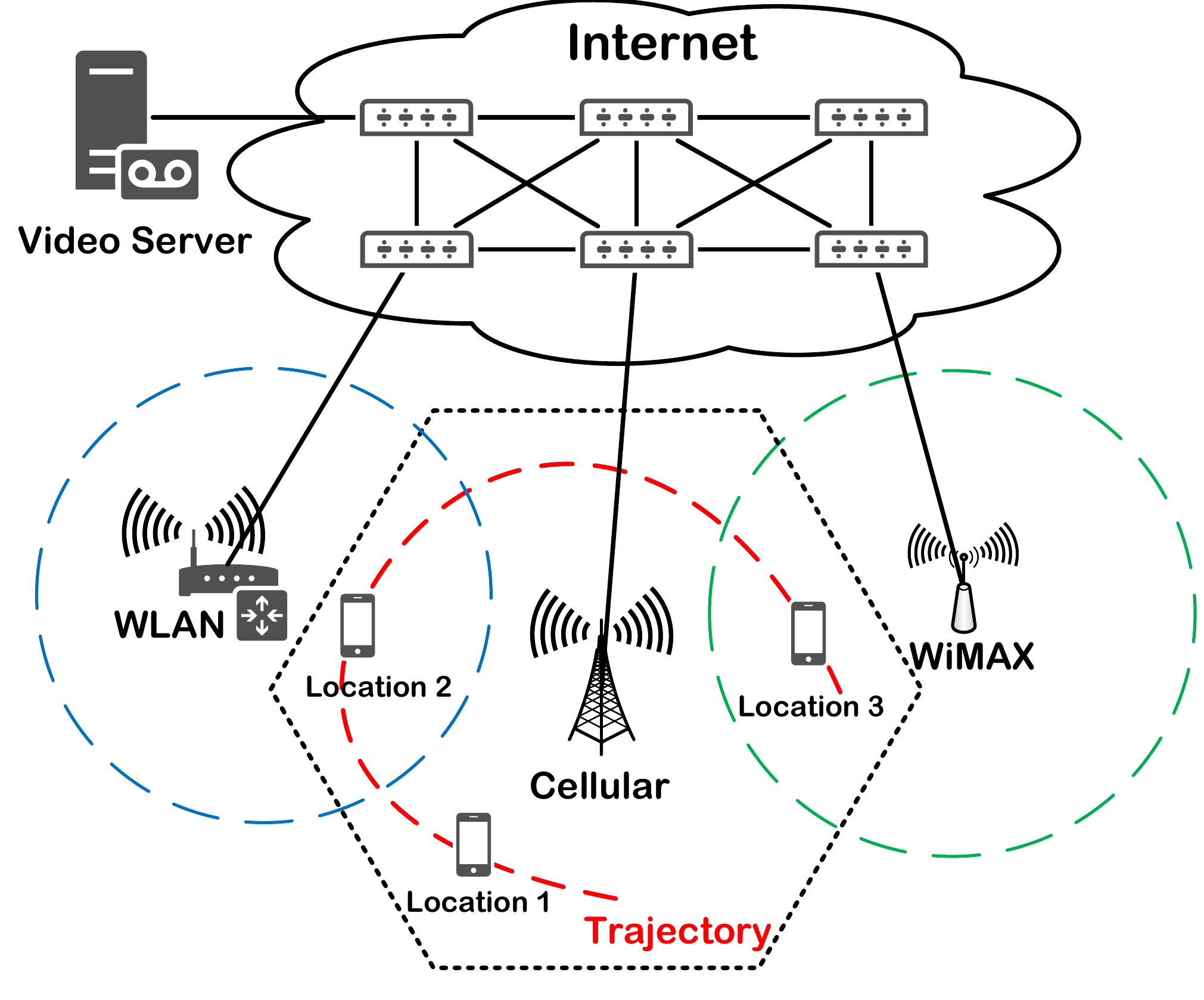}
 \caption{Illustration of a mobile video streaming service in a heterogeneous wireless network.}\label{fig1}
 \vspace{-5pt}
\end{figure}
In this study, we propose a Joint Data Allocation and Fountain Coding (JDAFC) approach to optimze JSCC to transfer mobile video streaming in heterogenous wireless networks. The key point of JDAFC is select appropriate wireless path and allocating proper data rate on each path. First, the data rate allocation solution is proposed to satisfy the realtime video transmission delay requirements. Second, the fountain code is employed to increase reliability of data transmission. Third, we proposed an effective flow rate allocating algorithm.
Specifically, the contributions of this paper can be summarized in the following:
\begin{itemize}
  \item We propose an end-to-end video transmission solution based on JSCC and path selection algorithm to improve video transfer quality in heterogeneous wireless network;
  \item We develop a mathematical model of JSCC to reduce the distortion of data transmission in multi-path wireless network.
  \item Perform emulations in Exata using real-time H.264 video.
\end{itemize}
The remainder of this paper is structured as follows. We discuss the related work in Section II. Section III depicts related model of the system and formulates the problem. In section IV, we describe the design of the proposed JDAFC in detail. The performance evaluation results are shown in Section V. Section VI gives the conclusion remarks.
\normalsize
\section{Related Work}
The related work to this study can be classified into two categories: (1) JSCC, and (2) transfer video in heterogeneous wireless networks. We will discuss on each topic respectively in this section.
\subsection{Joint Source Channel Coding}
Most study of the joint source channel coding in video transmission are concentrated on following items: 1) trying to figure out most appropriate video streaming allocation rate, e.g., [9]; 2) using different channel coding algorithm to satisfy the requirement, the Fountain [10], Turbo [11] and Reed-Solomon [12] codes; 3) considering the channel condition and design the video coding scheme to achieve the target, e.g., [13]. But the scenario of all this works are based on single path and waste multi-homed function of mobile devices.
\subsection{Transfer Video in Heterogeneous Wireless Networks}
Transfer video in heterogeneous wireless networks has been a hot area of research and [15-16] is a survey of current progression. The Encoded Multipath Streaming (EMS) [17] and Multi-Path LOss Tolerant (MPLOT) [18] are typical erasure code protocols for heterogenous network. The EMS scheme estimating network status (e.g., path loss rate) to allocate traffic loads over multiple paths. However, EMS was generally under the assumption that all the available paths could be beneficial for the transmission as in [14]. MPLOT is a transport protocol which aim is maximize the throughput of the upper-layer application. But MPLOT can not guarantee real-time video delivery as it does not address tight delay constraints.
\section{System Model and Problem Formulation}
The JDAFC system model is depicted in Fig. 2. The scenario we considered is a heterogenous wireless network with $\mathcal{R}$ wireless connections from a video server to a mobile device. This system involves three models: network model, end-to-end video quality model and fountain code model. Next subsection we simply describe each of them.
\begin{figure}[htbp]
\centering
 \includegraphics[width=0.5\textwidth,keepaspectratio]{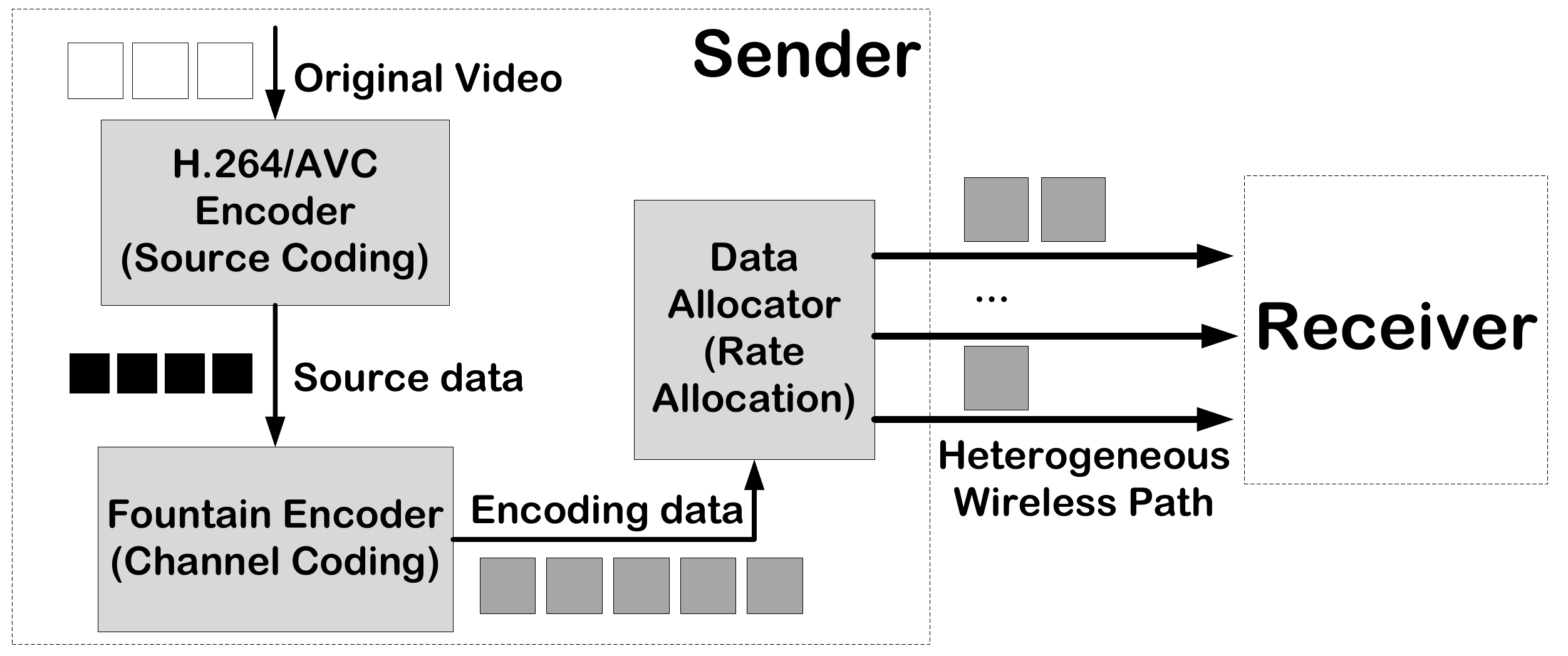}
 \caption{JDAFC system model with flow rate allocation over multi-path wireless networks.}\label{fig1}
 \vspace{-10pt}
\end{figure}
\subsection{Network Model}
In our network model, each physical path $P_r$ is described by the following metrics:
\begin{itemize}
  \item the available bandwidth $\mu_r$, expressed in the unit of Kbps.
  \item the propagation delay $t_r$, which includes the link delays of the wired and wireless .
  \item the average loss probability $\pi_B^r\in[0,1]$.
\end{itemize}
We employ continuous time Gilbert model to describe the burst loss behavior on physical path. The state $\mathcal{X}_{r}(t)$ assumes one of two values: $G$ (Good) or $B$ (Bad). $G$ means the packet is successfully delivered at time $t$. Otherwise, $B$ means the packet is lost.

We denote by $\pi^{r}_G$ and $\pi^{r}_B$ the stationary probabilities that $P_{i}$ is good or bad. Let $\xi^{r}_B$ and $\xi^{r}_G$ represent the transition probability from $G$ to $B$ and $B$ to $G$, respectively. In this paper, we adopt two system-dependent parameters to specify the continuous time Markov chain packet loss model: (1) the average loss rate $\pi_B^r$, and (2) the average loss burst length $1/\xi_B^r$. Then, we can have:
\begin{equation}
\begin{split}
\pi^{r}_{G} = \frac{\xi_B^r}{\xi_B^r+\xi_G^r},\text{ and }
\pi^{r}_{B} = \frac{\xi_G^r}{\xi_B^r+\xi_G^r}\cdot
\end{split}
\end{equation}

\subsection{Video Quality Model}
In this paper, a frame-level distortion model proposed in [19][20] is emplied to analyze the perceived quality of video streaming. In this study we only consider the constant bit rate (CBR) stream, and let $\gamma$ denote the video encoding rate. The video distortion is measured in units of mean squared error (MSE). If $N$ frames compose one GoP (Group of Pictures) and each frame is identified by a frame index $n$ ($1\leq n\leq N$). According to the affine model in [19], the total distortion($\mathrm{t}_n$) is mathematically expressed as follows.
\begin{equation}
\mathrm{t}_n=\mathrm{h}_n+\mathrm{d}_n,
\end{equation}
in which $\mathrm{t}_n$ represents the total distortion of frame $n$, $\mathrm{h}_n$ denotes the truncation distortion and $\mathrm{d}_n$ is the drifting distortion.
First, the truncation distortion $\mathrm{h}_n$ can be expressed as
\begin{equation}
\mathrm{h}_{n}={\theta'}_{n}+\tau_n\cdot\theta_n,\, 1\leq n\leq N,
\end{equation}
in which ${\theta'}_n$ denotes the full quality distortion of frame $n$, it means the quality degradation if the receiver get all the transmission packets of frame $n$. $\tau_n$ expresses the effective data loss rate of frame $n$, $\theta_{n}$ denotes the additional distortion introduced by dropping data packets in possibility.
Second, consider the GoP structure (\emph{i.e.}, IP...P) in this study, the previous frames within a GoP will impose drifting distortion on latter frames. We use the following equation to quantify the drifting distortion
\begin{equation}
\mathrm{d}_n={\eta}_n+\sum_{1< i< n}\mu_{n,i}\cdot \mathrm{h}_{i},1\leq n\leq N,
\end{equation}
%
where $\eta_{n}$ and $\mu_{n,i}$ are the parameters estimated from real video data. And $\mathrm{d}_n$ will be when $n$ equals $0$. We employ the global evaluation method proposed by [20] to ensure the low-complexity of the parameter estimation process for online operation.
\subsection{Fountain Code}
Figure 3 illustrates the application of fountain coding in video streaming transmission procedure. In this work, the GoP-level fountain coding is employed for the data protection. The coding element of fountain code is symbol which is 'source block' essentially, so the source stream is encoded to video stream and then divided into source symbols. These symbols are input of fountain encoder.

Then we need to XOR some source symbols to generate an encoding symbol, the number of XOR source symbol is defined as the degree (n degree), which is one of the most important parameter in fountain code. The domain of n degree is from 1 to k when the symbol counts is k. To determine the n degree, the probability distribution \{$p_1$,$p_2$,...,$p_k$\} is firstly allocated to \{1,2,...,k\} values according to a pre-determined probability function. Then we set to n degree a random number is generated using the above probability distribution \{$p_1$,$p_2$,...,$p_k$\}. Now, the n degree source symbols are uniform randomly selected and then an encoding symbol is generated by performing bitwise XOR operations on the selected n degree source symbols.

Since an encoding matrix realization can be characterized by the seed number of the pseudo-random number generator [24], a receiver can easily reconstruct source symbols from encoding symbols if sender and receiver share the same seed number. Hence, the required side information for the fountain decoding is negligible. At the receiver, all the source symbols can be recovered if a sufficient number of encoding symbols are available even though some encoding symbols are lost. In general, the number of encoding symbols required for successful fountain decoding is calculated by
\begin{equation}
{m}{'}=(1+{\psi}(m))\cdot{m},
\end{equation}
where ${\psi}(m)$ is the symbol overhead according to the number of source symbols, which is a function of k. The above equation means that the number of received encoding symbols must be slightly larger than k to reconstruct the source symbols successfully. In the FEC scheme including fountain code, the code rate (r) plays an important role because it determines the amount of redundant data for error protection. That is
\begin{equation}
{r}=m/n,
\end{equation}
where $n$ is the number of encoding symbols. Accordingly, the control parameters of the fountain encoding process are $m$ and $r$. Both of them have to be determined before the fountain encoding operation takes place.

%
\begin{figure}[htbp]
\centering
 \includegraphics[width=0.5\textwidth,keepaspectratio]{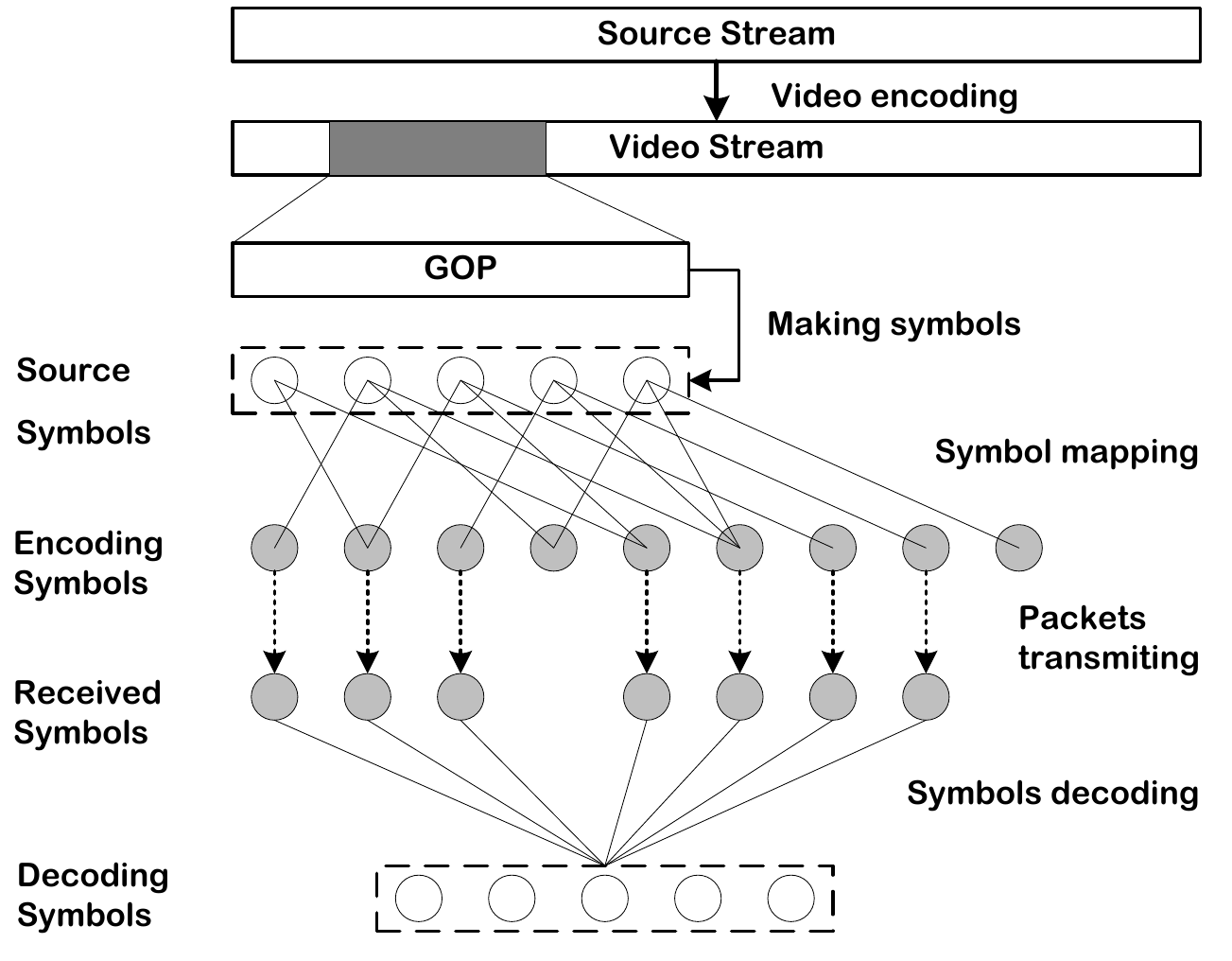}
 \caption{GoP level Fountain coding illustration.}\label{fig1}
 \vspace{-10pt}
\end{figure}

\section{Design of Joint Data Allocation and Fountain Coding System}
We present the overall design of the proposed JDAFC and separately introduce the main components in this section. Fig. 4 shows the system framework. Both the server and client contain the implemented components to solve the problem, server side includes: (1) flow rate allocation, (2) path selector. Particularly, the data allocator component is response to allocate encoded fountain symbols into available heterogenous wireless pathes.

Client side contains an information feedback component which send network status information back to server side. And the other parts are normal
components of video client. Packets are received by packet receiver, then the video frames will be stored in the playback buffer after the fountain code decoding process. The data flow reassemble components reorder the decoding symbols and obtain correct video frames. Next, we will describe the key components in the system design and their major functions.

%

\begin{figure}[htbp]
\vspace{-5pt}
\centering
 \includegraphics[width=0.5\textwidth,keepaspectratio]{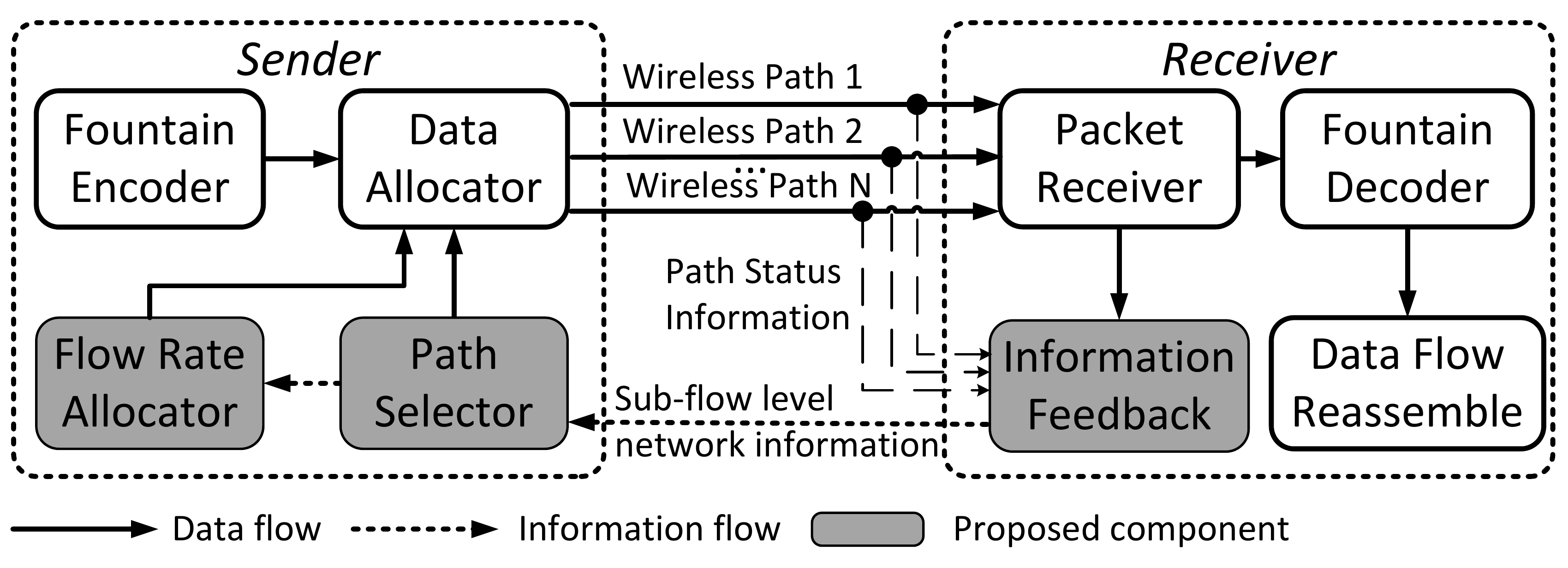}
 \caption{Overall design of the proposed JDAFC, which consists of working components at the server and the client side}\label{7}
 \vspace{-10pt}
\end{figure}
\subsection{Path Selector}
In this study we employ an system availability model developed in [23]. This solution minimize the sum of total distortion by allocating video frames according to 'reliability level' of the communication path, this reliability parameter quantified the path availability. Therefore, the higher reliability path get higher priority to transfer video frames in the path selection algorithm. Due to the path asymmetry of heterogeneous networks, the load imbalance problems may always occur [25][26]. The mitigate severe imbalance in heterogenous paths, we employ another load imbalance parameter to indicate whether the path is overloaded and we can infer in [25].Then the video data is allocated to available paths. And on the other hand, the video frames with the highest distortion will be dropped to conserve bandwidth, and thus improve the overall streaming quality.
\subsection{Flow Rate Allocator}
The goal of the flow rate allocator is allocate optimal flow rate of the flow to all the candidates path so as to minimize total video distortion. In order to optimize the flow rate, we first get an available wireless path list sorted by their 'loss-free' bandwidth, which could effectively indicate the path quality[18], then we can allocate proper flow to each path in proportion according to the weight of it in the path list. This is a simple but effective calculate algorithm.
\subsection{Information Feedback}
The available bandwidth, channel loss rate and propagation delay are the most important parameters are especially important properties for a high quality video streaming service. Information feedback component is used to collect this necessary information from all the wireless paths and transfer it back to server side. Using feedback path information to estimate channel status has been attracting research attentions for years. It is very important to identify the physical characteristics of each wireless path so as to utilize channel resources efficiently of heterogenous wireless network.
\section{Performance Evaluation}
In this section, we conduct a experimental environment to evaluate performance of proposed JDAFC solution, and then compare it with existing schemes.  First, we will describe the emulation methodology that contains emulation scenario, emulation setup and performance metrics.
\subsection{Emulation Methodology}
\subsubsection{Emulation Scenario}
\begin{figure}[htbp]
\vspace{-5pt}
\centering
 \includegraphics[width=0.4\textwidth,keepaspectratio]{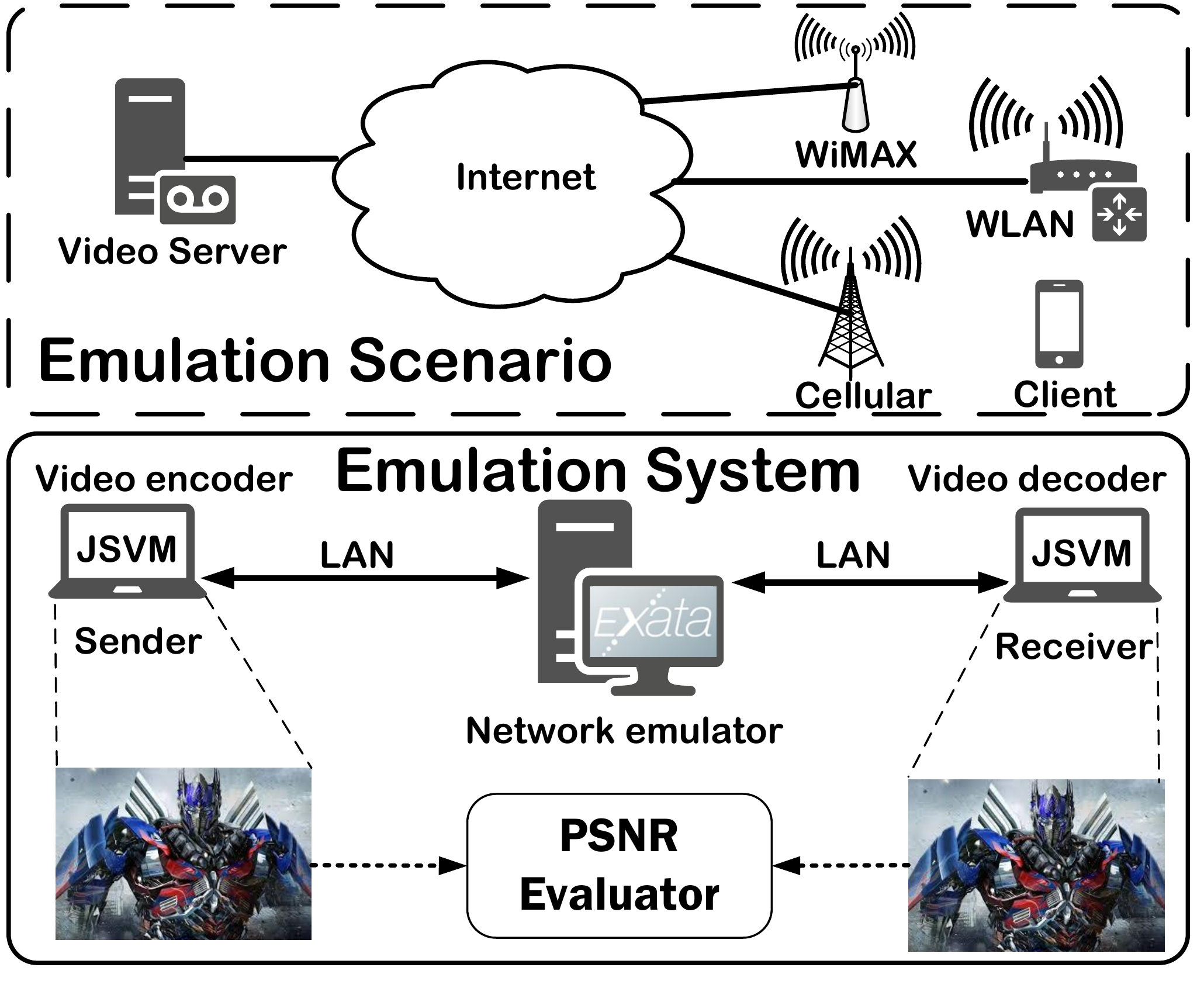}
 \caption{System architecture for performance evaluation.}\label{7}
 \vspace{-10pt}
\end{figure}
We conduct all the simulation experiment in a mobile device scenario as shown in Fig. 5. We establish four unique combinations to achieve reliable experiment result. The four combinations represent dynamically select various access options(e.g. LTE, WiMAX and WLAN) for the mobile user in the integrated heterogeneous wireless networks, the combinations are listed in table 1. The mobile client requests to the server through a wireless interface and constructs the connection whenever it moves in the coverage. The moving speed of the client is set to be $2$ m/s in all the experiment.
\begin{table}[htbp]
\scriptsize
\setlength{\tabcolsep}{1pt}
\caption{Heterogenous Wireless Networks Combinations}
\centering
 \begin{tabular}{|c|c|}
    \hline
   Index & Combinations \\
    \hline
    \hline
    1 & LTE+WiMAX (All the time)\\
    \hline
    2 & LTE+WLAN (All the time)\\
    \hline
    3 & LTE+WiMAX+WLAN (All the time)\\
    \hline
    4 & LTE $\rightarrow$ LTE+WiMAX $\rightarrow$ LTE+WiMAX+WLAN $\rightarrow$ LTE+WLAN $\rightarrow$ LTE\\
    \hline
  \end{tabular}
\end{table}
For the confidence results, we repeat each exprement with different video sequences more than $5$ times and obtain the average results with a $95\%$ confidence interval.
\subsubsection{Emulation Setup}
We employ the Exata and JSVM applications as the network emulator and video codec respectively. The constitution of the evaluation system is depict in Fig. 5 and the main configurations are set as follows.
\begin{itemize}
    \item Exata 2.1 is adopt as the network emulator. The server and client are corresponding to real computers, which are connected to the emulation server through the Exata connection Manager. The IEEE 802.11b is adopted as the WLAN protocol. And heterogenous wireless networks configurations can refer to [4-5][22].

    \item JSVM 9.18 is adopted as the H.264/SVC video encoder software. The encoding rate of generated video is $30$ frames per second and one GoP composed by eight frames. The test video sequences are \emph{Sky, Ducks, Tree, and Stockholm} in QCIF (Quarter Common Interchange Format) format with $300$ frames. We join $10$ video sequences together and get a $3000$ frame-long sequence so as to obtain statistically meaningful results. The loss requirement and delay constraint  are set to $1\%$ and $250$ ms respectively.
\end{itemize}

\subsubsection{Comparing Schemes}
We compare the performance of JDAFC with the following schemes for video delivery in heterogeneous wireless networks.
\begin{itemize}
    \item EVPCS [6]. [6] proposed an 'End-to-End Virtual Path Construction System' based on fountain code aims at exploiting path diversity over heterogeneous wireless networks, in this paper we name it EVPCS.
    \item JFSS [14]. The Joint FEC and Scheduling Scheme (JFSS) computes the optimal source and FEC rate for scalable video over multipath networks.
    \item DMP [21]. The Dynamic MPath-streaming (DMP) utilizes multiple paths by maintaining a TCP connection on each path. The sender only use the best TCP connection sending data at any time until the connection is blocked. Another available TCP connection will then gain the access to the sender queue and continue sending data.
\end{itemize}

\subsubsection{Performance Metric}
We employ some frequently-used performance metrics to evaluate the proposed approach against the above comparing approaches:
\begin{itemize}
  \item PSNR. A standard metric of video quality, which is a function of the mean square error between the original and the received video frames.
  \item Average end-to-end delay. A metric which consists of delay in the network and the resequencing time at the client of a video frame.
  \item Effective loss rate. The effective loss rate is used to testify the competing models in mitigating the packet loss.
\end{itemize}

\subsection{Evaluation Results}
\subsubsection{PSNR}
\begin{figure}[htbp]
        \centering
         \includegraphics[width=0.4\textwidth,keepaspectratio]{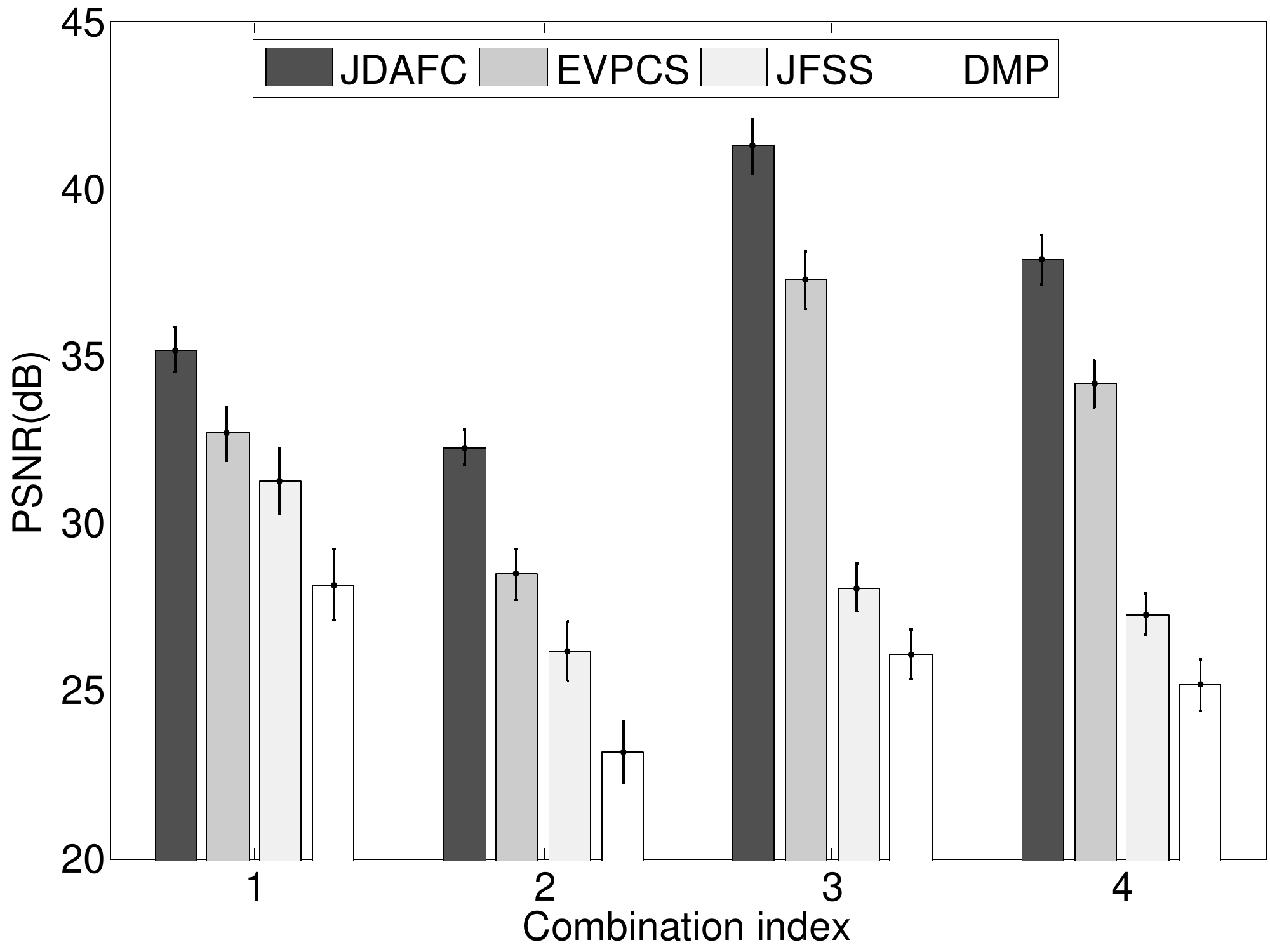}
         \caption{Average PSNR values and variances under difference scenarios.}\label{8}
\end{figure}
Figure 6 shows JDAFC achieves higher PSNR values and lower variations than the other comparing models. As the WLAN is less stable than the WiMAX, causing combination $2$ get lower average video PSNR than combination $1$. The results verify the concrete example in Figure 1. Besides, the superiority of JDAFC and EVPCS over the other two models is larger in combination $3$ and $4$ as more wireless access networks are available. Fig. 7 also depicts the mean values and standard deviations (Stddev) of combination $4$ so as to have a microscopic view of the results.


\begin{figure}[htbp]
        \centering
         \includegraphics[width=0.4\textwidth,keepaspectratio]{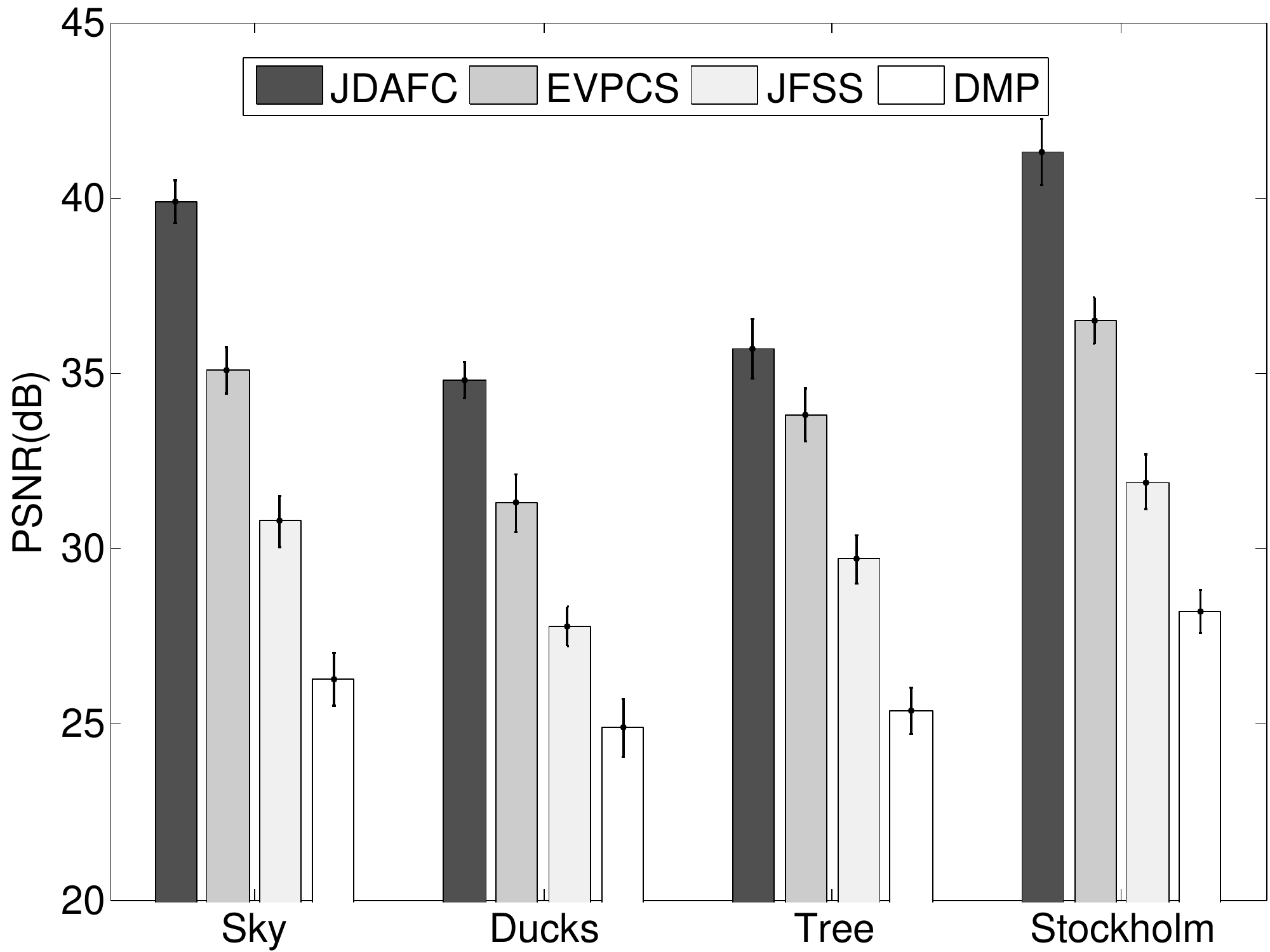}
         \caption{Average PSNR for different compared models of combination 4.}\label{8}
\end{figure}

\subsubsection{Average end-to-end delay}
Fig. 8(a) shows the cumulative distribution function (CDF) of the end-to-end video frame delay in a single experiment. We can see compare to other schemes, that the per frame delay of JDAFC is significantly lower. Figure 8(b) represents the ratio of video frames past the decoding deadline of $200$ ms because each video frame is associated with a decoding deadline in real-time applications. Figure 9 plots the average end-to-end delays as well as the confidence intervals, JDAFC achieves the lowest delay of all the competing models.
\begin{figure}[htbp]
\centering
    \begin{minipage}[t]{0.49\linewidth}
     \centering
    \includegraphics[width=1\textwidth,keepaspectratio]{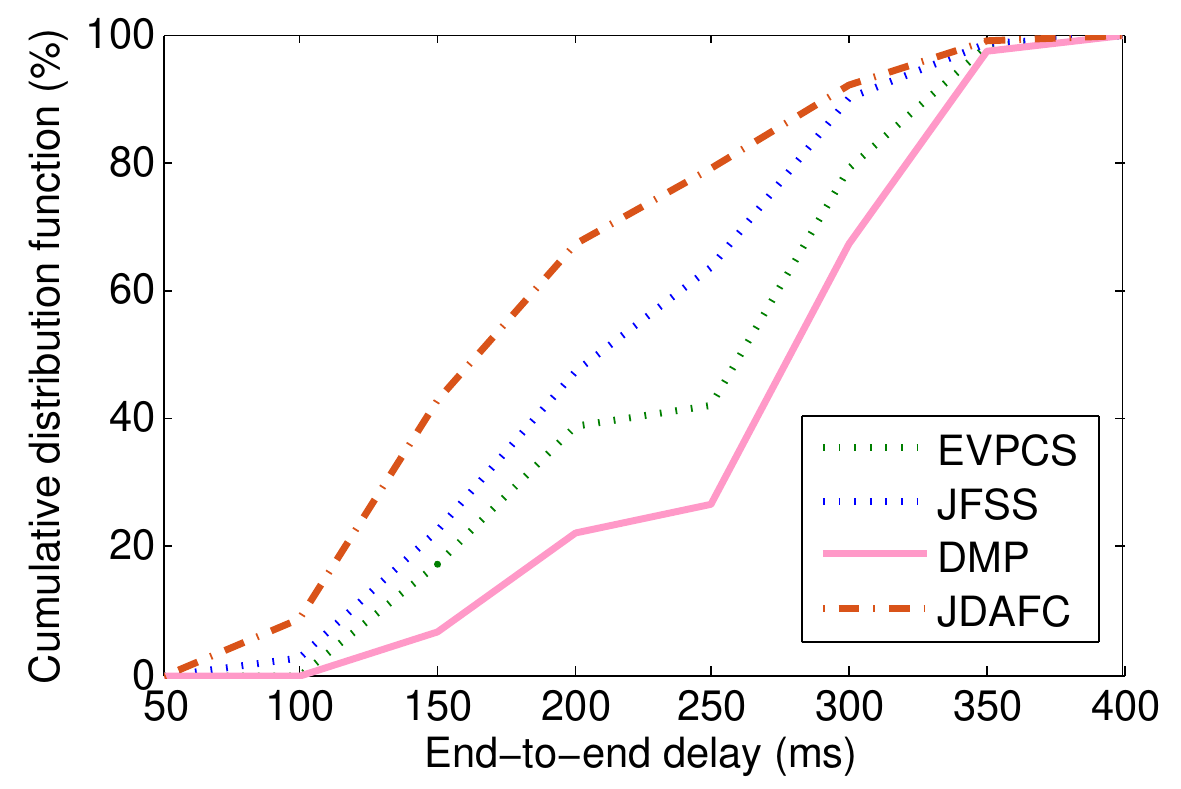}\\
        \centering
     \label{fig:side:a}
    \end{minipage}
     \begin{minipage}[t]{0.49\linewidth}
     \centering
     \includegraphics[width=1\textwidth,keepaspectratio]{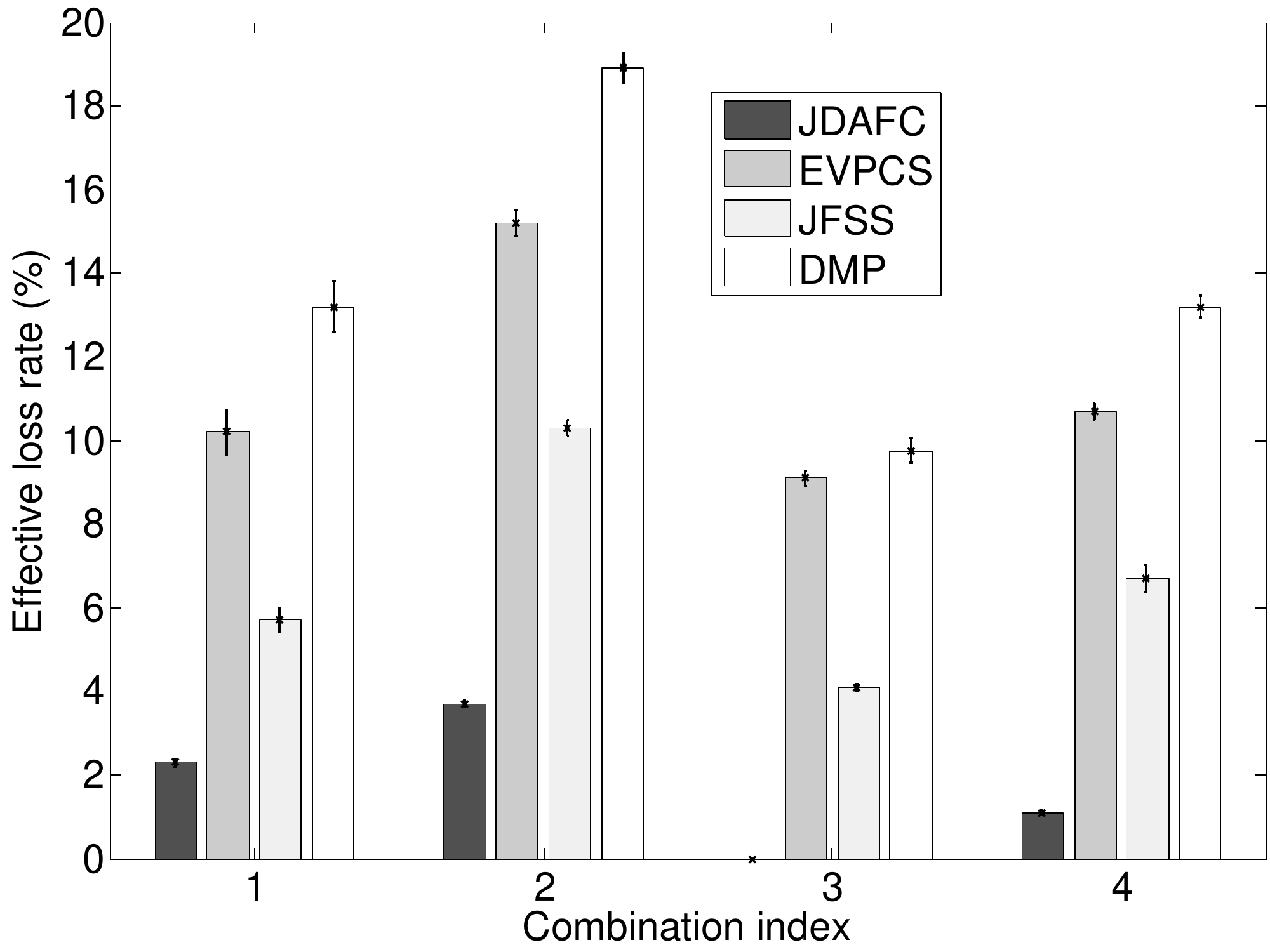}\\
        \centering
     \label{fig:side:a}
    \end{minipage}
\caption{(a) cumulative distribution function, (b) ratio of video frames past decoding deadline of $200$ ms.}\label{8}
\end{figure}
\subsubsection{Effective loss rate}

Figure 10 depicts the effective loss rates of all the competing schemes of different network combinations. We can significantly obtain that JDAFC outperforms the competing schemes as it considering both the loss and delay requirements.
\begin{figure}[htbp]
        \centering
         \includegraphics[width=0.4\textwidth,keepaspectratio]{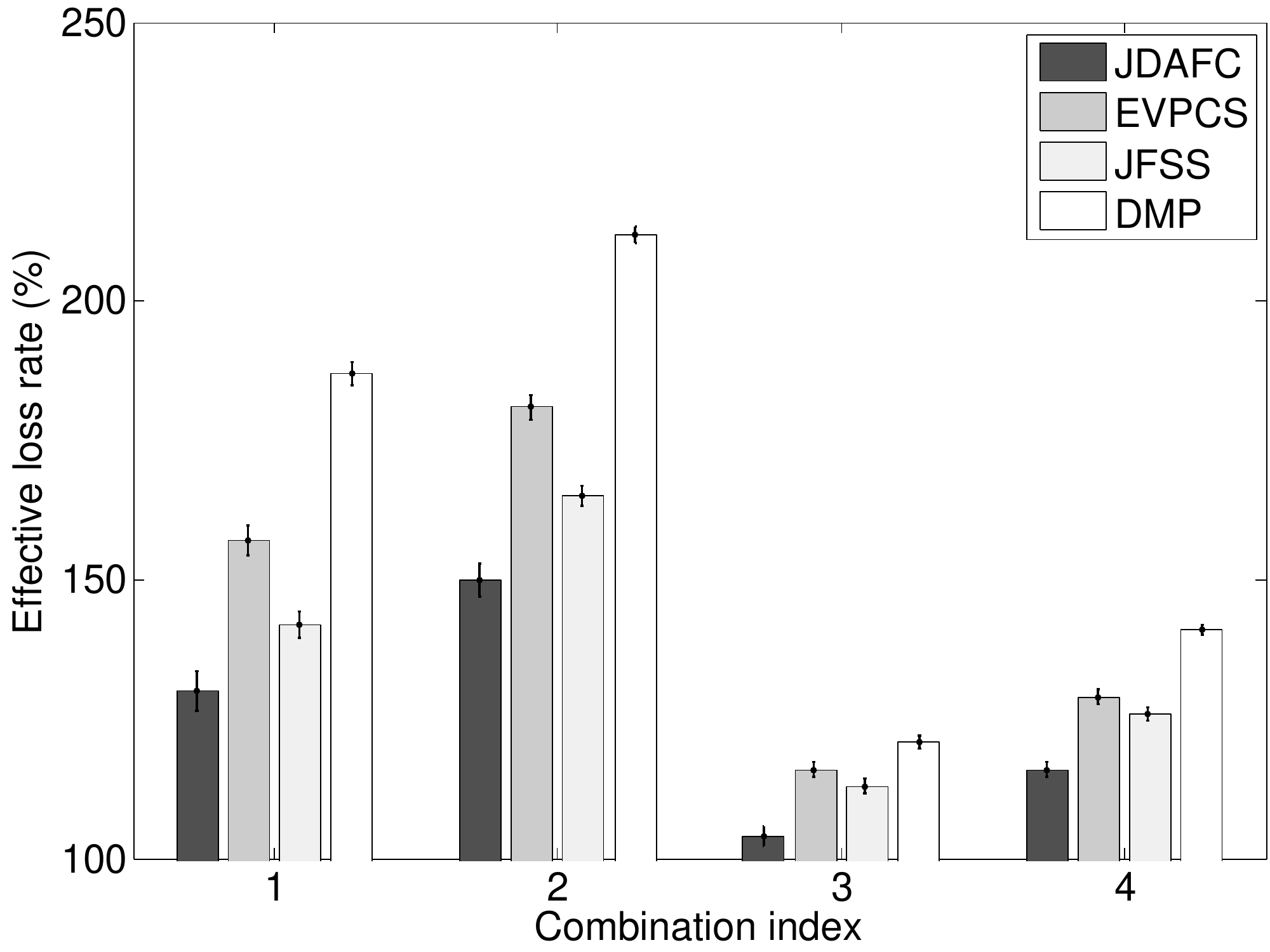}
         \caption{The average end-to-end delays of all the schemes.}\label{8}
\end{figure}
\begin{figure}[htbp]
        \centering
         \includegraphics[width=0.4\textwidth,keepaspectratio]{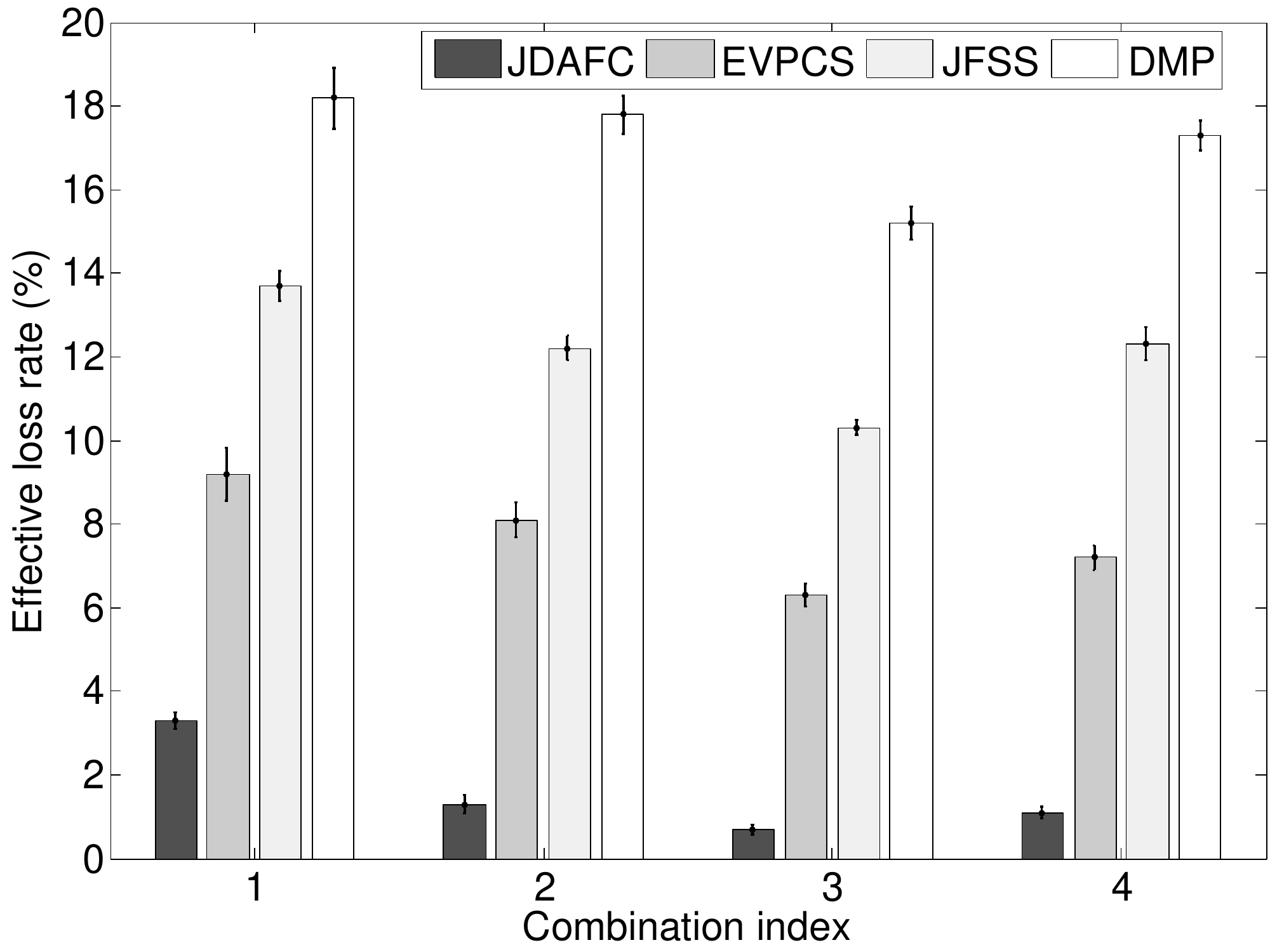}
         \caption{The effective loss rates of all the models.}\label{8}
\end{figure}

\section{Conclusion and Discussion}
In this paper we present a Joint Data Allocation and Fountain Coding(JDAFC) approach for mobile video delivery in heterogeneous wireless networks. We proposed solutions for available path selection, flow rate allocation and network information feedback. The simulation results depict that the proposed JDAFC possess capability to dynamically select the appropriate wireless access networks out of all candidates and obviously improve the video PSNR. As future work, we will consider:  (1) design a seamless vertical handoff algorithm for optimal-quality video in the integrated WLAN, WiMAX and Cellular networks. (2) include an optimal path interleaving mechanism with the JDAFC for overcoming the burst loss.

\end{document}